\documentclass[12pt]{article}
\usepackage{amssymb}
\usepackage{graphicx}
\usepackage{epsf}
\usepackage{lscape}
\usepackage{epstopdf}
\usepackage[cp866]{inputenc}
\usepackage[T2A]{fontenc}
\usepackage[english]{babel}
\usepackage{amsmath}
\begin{document}

\title{\large{\bf   Asymptotic normalization coefficients for   ${\rm {^7Be}}+p\to{\rm{^8B}}$ from the peripheral         ${\rm{^7Be}}$($d$,$n$)${\rm{^8B}}$ reaction and their astrophysical application    }}
\author{O. Tojiboev$^{1)}$,   R. Yarmukhamedov$^{1)}$\thanks{Corresponding author, E-mail:
rakhim@inp.uz}\,, S. V. Artemov$^{1)}$, S. B. Sakuta$^{2)}$ }

\maketitle{\it
$^{1)}$Institute of Nuclear Physics,  Tashkent 100214, Uzbekistan\\
$^{2)}$National Research Center ``Kurchatov Institute'', Moscow 123182, Russia\\}

\begin{abstract}
 The     proton transfer
  ${\rm{^7Be}}$($d$,$n$)${\rm{^8B}}$ reaction at the  energy of 4.5 MeV (c.m.)   has been analysed within the modified  DWBA.  New estimates and their uncertainties are obtained for
values of  the asymptotic normalization coefficients for
 $ p+{\rm{^7Be}}\to{\rm{^8B}}$,    the astrophysical $S_{17}(0)$ factor and the s-wave $p$+${\rm{^7Be}}$ scattering lengths.
 \end{abstract}


 \vspace{1.0cm}
 \begin{center}
 {\bf I. INTRODUCTION}
 \end{center}
\vspace{1.0cm}

\hspace{0.6cm} The   radiative capture ${\rm{^7Be}}$($p$,$\gamma$)${\rm{^8B}}$
 reaction rate given in terms of the extremely-low energy astrophysical $S$ factor is one of the main input data  for   calculations of   the    solar-neutrino flux [1--6]. At the  stellar temperature $T_6\sim $ 15 K,   this rate  determines    how much  the  ${\rm{^7Be}}$  and ${\rm{^8B}}$  branches of   the   $pp$ chain   contribute to the  hydrogen burning. In the standard solar model, the predicted flux of  neutrinos is   determined   by the relation  \cite{Bah1969}
 \begin{equation}
\label{eq1}
\phi_{\nu}\sim
\tilde{S}_{11}^{-2.5}\tilde{S_{33}}^{-0.3}\tilde{S_{34}}^{0.8}
[1+3.47\tilde{S_{17}}\tilde{\tau}_{e7}^{-1}].
\end{equation}
Here $\tilde{S_{ij}}=S_{ij}(E)/S_{ij}(0)$  and  $\tilde{\tau_{e7}}=\tau_{e7}(E)/\tau_{e7}(0)$, where  $S_{ij}(E)$ is   the extremely-low energy astrophysical $S$ factor  for  the   reactions of the $pp$ chain  induced  by collisions  of   the nuclei with  mass numbers $i$ and $j$, $\tau_{e7}(E)$  is   the ${\rm{^7Be}}$     lifetime with respect  to the  electron capture  ${\rm{^7Be}}+e^-\to{\rm{^7Li}}+\nu_e$ reaction and  $E$ is  the relative kinetic  energy  of the  colliding particles.    It is seen that the flux of neutrinos depends noticeably on the $\beta$-decay of   ${\rm{^8B}}$ determined in turn  by   the accuracy of the   astrophysical $S$ factors of the  ${\rm{^7Be}}$($p$,$\gamma$)${\rm{^8B}}$ reaction at so far experimentally inaccessible solar energies ($E\lesssim$  25 keV), including   $E$=0.
 
    It is known  that the uncertainty  in   extrapolation of the astrophysical $S$ factors   to the Gamow energy $E_G$=18  keV      obtained at the stellar temperature  $T_6$=15 K (Sun)  \cite {Rolf88}   affects significantly    the predicted flux of solar     ${\rm {^8B}}$  neutrinos  \cite{Bah98,Adel2011}.

Despite    the impressive improvements in our understanding   the nuclear-astrophysical  ${\rm{^7Be}}$($p$, $\gamma$)${\rm{^8B}}$  reaction   made  in  a   period of ten consecutive years,  some ambiguities connected with both the extrapolation of the astrophysical $S$ factors        to the solar energy region  and the theoretical prediction for    $S_{17}(E)$ still exist and     can  influence the prediction of
the standard solar model  \cite{Bah98}.  For example, the   analysis   of  the      precisely measured      astrophysical  $S$ factors   for the ${\rm{^7Be}}$($p$, $\gamma$)${\rm{^8B}}$  reaction    and  their extrapolation  performed  by different authors  from the observed  energy regions to extremely low experimentally inaccessible energies     gives      values of   $S_{17}(0)$  with  a spread exceeding noticeably the experimental ones (see the recent  reviews  \cite{Adel2011,Yarm2013} and references therein as well as  Ref. \cite{Igam08}).   As regards   the theoretical microscopic  calculations of   $S_{17}(0)$,   they also show  considerable spread connected with the method used (see,  for example,  Refs.  [11--13] and  the available   references therein).  It should be noted that        a  considerable  sensitivity of   the calculated  value of  $S_{17}(0)$ is observed in [11--13]         to   the used effective nucleon-nucleon potential.  Moreover,  as it is evident  from these works,   a correlation has been revealed  between the   asymptotic normalization   coefficients (ANC) calculated for   $p+{\rm{^7Be}}\to{\rm{ ^8B}}$,   which determine the amplitude of the tail of the radial  ${\rm
{^8B}}$ nucleus bound    wave function in the (${\rm
{^7Be}}$+$p$)-channel  \cite{Blokh77}, and  the calculated  $S_{17}(0)$.

Taking into  account this fact,   in the last years several works have been  performed    to determine  the    ANC     for  $p+{\rm{^7Be}}\to{\rm{ ^8B}}$ and the         $S_{17}(0)$     with  an accuracy as high as possible.     In  \cite{Igam08},    the ``indirectly determined''   value of   ANCs  for  ${\rm {^7Be}}+p\to{\rm{^8B}}$      were obtained  by means of   the   analysis of the precisely measured   experimental astrophysical  $S$ factors ($S_{17}^{exp}(E)$) performed within the  modified two-body potential  approach (MTBPA)   \cite{Igam07} where     $S_{17}(E)$ for the direct radiative capture ${\rm{^7Be}}$($p$, $\gamma$)${\rm{^8B}}$  reaction  is expressed in terms of ANC for $p+{\rm{^7Be}}\to{\rm{ ^8B}}$.   Then  the   derived   ANC     was  used for  extrapolation of  the corresponding astrophysical $S$ factors to solar energies.   In \cite{Tab06} (see also  \cite{Azh1999a,Azh1999b}),  the ANC values   for  $p+{\rm{^7Be}}\to{\rm{ ^8B}}$ were obtained from  analysis of  the peripheral proton transfer ${\rm {^{10}B}}$(${\rm{^7Be}}$,${\rm{^8B}}$)${\rm{^9Be}}$ and ${\rm{^{14}N}}$(${\rm{^7Be}}$,${\rm{^8B}}$)${\rm{^{13}C}}$ reactions. The analysis  was  performed within the modified DWBA restricting by  the first order of the perturbation theory over the Coulomb polarization potential ($\Delta V_f^C$) in the transition operator and assuming that its contribution to the total transition operator is small. But,    the ANC values proposed in  \cite{Tab06} may not have sufficient accuracy because of  the aforementioned assumption made for the transition operators \cite{Igam072,Ogata2006}. Nevertheless, in \cite{Tab06} the obtained  ANC values were then  used for the  estimation of    $S_{17}(0)$, which  also differs  noticeably from  that   recommended    in     \cite{Adel2011,Igam08, Jun03}.   As  is  noted in    \cite{Igam08,Ogata2006},  one of the main reasons of the observed  discrepancy  is     connected    with       underestimated values of the ANCs  suggested  in \cite{Tab06} as compared with  that obtained in   \cite{Igam08,Ogata2006}.

This disagreement  initiated   a number of the new   measurements of other kinds of peripheral nuclear processes   to  obtain an additional information about  the ANC values  for  ${\rm {^7Be}}+p\to{\rm{^8B}}$ and their  astrophysical application.  Particularity,   the analysis of   the experimental differential cross sections for  the   peripheral proton transfer  ${\rm{^7Be}}$($d$,$n$)${\rm{^8B}}$ reaction, measured in  inverse kinematics   at the energies 5.8 \cite{WL1998} and 4.5 MeV \cite{Das2006} in the center of mass system (denoted by $E_i$ everywhere below),  was performed  in   \cite{WL1998,Ogata2003} and \cite{Das2006}, respectively. In \cite{WL1998,Das2006} and \cite{Ogata2003},  the analysis was carried out within the standard   DWBA and   the three-body approach with the effects of breakup of the deuteron  in the field of the target   treated by means of continuum-discretized coupled-channel  method (CDCCM). There the calculated    cross sections are expressed  in terms of the spectroscopic factors for   the ${\rm{^8B}} $  nucleus in the (${\rm{^7Be}}+p$)-configuration. Besides, in \cite{WL1998},  for the transferred proton only the 1$p_{3/2}$ configuration was only taken into account.  However, the extracted in \cite{WL1998,Ogata2003} and \cite{Das2006}    spectroscopic factors,   used for obtaining the ANC values,    may not be accurate enough.  This is connected with the fact that   the  calculated  single particle   cross sections, as  will be shown below,   become very sensitive   to the adopted  values of the geometric parameters of the Woods-Saxon potential  used    for  calculation  of the bound ${\rm{^8B}}$[=(${\rm{^7Be}}+p$)] state wave function   \cite{Gon82}.  The results of the analysis of the experimental differential cross sections \cite{WL1998} of  the ${\rm{^7Be}}$($d$,$n$)${\rm{^8B}}$ reaction     obtained in   \cite{Toj2014} within the modified DWBA [27--29], where the differential cross section is expressed in terms of   the ANC but not in terms of  the spectroscopic factors,  should also  be noted.
 In \cite{Toj2014},  the ANC values for ${\rm {^7Be}}+p\to{\rm{^8B}}$ were  obtained  by using the optical potentials recommended in \cite{WL1998} for the entrance and exit channels, but    contribution of the compound nucleus evaluated in \cite{WL1998} and the coupled channels effects were not taken into account. So, it should be noted that, firstly, the experimental data of \cite{WL1998} in the main peak region of the angular distribution  have  fairly large errors not only  in the absolute values of the differential cross sections  but  also  in the angle resolution.
    Secondly, as    is shown in \cite{Das2006},  calculations with the optical parameters of \cite{WL1998} for the entrance channel,   which are also used in \cite{Toj2014},  reproduce   the corresponding experimental  scattering data only in a narrow angular interval of  the forward hemisphere. Therefore, the ANC values of \cite{Toj2014} also may   not have enough accuracy for the astrophysical application. In this connection, it is of great interest to apply the modified DWBA for the analysis of the experimental data of the ${\rm{^7Be}}$($d$,$n$)${\rm{^8B}}$ reaction measured in \cite{Das2006} since they are more  peripheral  and more precise than  the data  of  \cite{WL1998}.

In the present  work, the reanalysis of the  experimental differential  cross sections \cite{Das2006}   for the    ${\rm{^7Be}}$($d$,$n$)${\rm{^8B}}$    reaction at energy $E_i$=4.5 MeV   is performed within the modified DWBA [27--29]  to obtain  the  ``indirectly     determined''    values  of  the ANC    for ${\rm {^7Be}}+p\to{\rm{^8B}}$, which were then  used  for  estimation of  the    $S_{17}(0)$.  Here,  we quantitatively show that  the ${\rm{^7Be}}$($d$,$n$)${\rm{^8B}}$  reaction measured in \cite{Das2006}  is   practically  peripheral  in     the main peak region of  the  angular distribution.    Therefore,   the ambiguities, which are  connected both with the variation  of  the geometric parameters (radius $r_o$,   diffuseness $a$) of  the Woods-Saxon potential used  for calculating of  the bound state wave functions and with   choice of  a set of the optical parameters,   are reduced  to the  physically acceptable limit, being within the errors  of  the  analyzed experimental  differential cross sections.

\vspace{1.0cm}

\begin{center}
{\bf II.  ANALYSIS OF  THE  ${\rm{^7Be}}(d,n){\rm{^8B}}$  REACTION}
\end{center}

\begin{center}
{\bf A. The   basic formulas of the modified DWBA} 
\end{center}

 Here   we  present  only the main idea and the essential  formulas of the MDWBA \cite{Ar96,Mukh97} specialized  for the  ${\rm{^7Be}}$($d$,$n$)${\rm{^8B}}$ reaction.
 
 Let us write $l_B$($j_B$) for the orbital (total) angular momentum of the proton in the ${\rm{^8B}}$ nucleus and $l_d$($j_d$) for the proton in the deuteron.  For the  ${\rm{^7Be}}$($d$,$n$)${\rm{^8B}}$ reaction
the value of $l_B$ is taken to be equal to 1  and the values of $j_B$ are taken to be equal to 1/2 and 3/2, while the values $l_d$ and  $j_d$ are taken equal to 0 and 1/2 for $s$-wave and  to 2 and 3/2 for $d$-wave, respectively.
 
The main contribution to  the  full transition operator  in the amplitude   (at least near the main peak region in the angular distribution) comes from    the proton-neutron nuclear $V_{np}^N$ potential \cite{Igam072,Austern1966,LOLA} since $\Delta V_f^C$=0 in the  full transition operator  because of the presence of a  neutron in the final state. Then the single particle DWBA cross section    can be   calculated  within the finite range DWBA in the ``post''-approximation with   the  DWUCK5 code. But if the reaction under consideration is peripheral, the influence of the ambiguity  of  the optical model parameters on the calculated single particle DWBA cross section should be   small, at least   not exceeding  the experimental errors. In this case, the largest uncertainty in the  single particle DWBA cross section comes from  its strong  dependence  on the geometric parameters (radius $r_0$ and diffuseness $a$) of the Woods--Saxon potential used 
for calculating the radial wave function of  the single-particle ${\rm{^8B}}$[=(${\rm{^7Be}}+p$)] bound state wave function. 
  It is known that  this dependence enters  the single particle DWBA cross section mainly through the corresponding single-particle ANC \cite{Gon82}.  It will be demonstrated below that the  reaction ${\rm{^7Be}}$($d$,$n$)${\rm{^8B}}$  at $E_i$=4.5 MeV is peripheral, so the single-particle ANCs for the shell model bound wave functions of the residual nucleus 8B, entering the calculated single particle DWBA cross section, are the free parameters. 
    
  Only  $s$ wave is  taken into account for the deuteron wave function in our calculations. This approximation     is justified  by  the fact that the reaction under consideration has  the peripheral character at least in the main peak region of the angular distribution. Therefore,  in this angle region the dominant  contribution to the DWBA cross sections comes from the surface and outer regions of the colliding nuclei.  In   this interaction region,   contribution of   the $d$ wave component  to the  deuteron  wave     function is strongly suppressed  since the   amplitude of its ``tail'' determined by the corresponding   ANC  of the $d$ wave is very small ($\sim$10$^{-2}$ fm$^{-1/2}$)   as compared  to that for the    $s$ wave    \cite{Blokh77,MB}.  

Then, according to  [27--29], within the modified DWBA, we can write the  differential  cross section in the form
\begin{equation}
\label{eq2}
 \frac{d\sigma}{d\Omega} =C^2_{B;\,3/2}[{\cal {R}}_{3/2}(E_i,\theta;\,b_{B;3/2})+\lambda{\cal {R}}_{1/2}(E_i,\theta;\,b_{B;1/2})],
\end{equation}
\begin{equation}
\label{eq3}
{\cal {R}}_{j_B}(E_i,\theta;\,b_{B;\,j_B})=\left(\frac{C_{d;\,1/2}}{b_{d;\,1/2}}\right)^2
\frac{\sigma^{\rm
{DW}}_{j_B}(E_i,\theta;\,b_{B;\,j_B})}
{b_{B;\,j_B}^2} ,
\end{equation}
where $C_{d;\,j_d}$ and  $C_{B;\,j_B}$ are  the ANCs for $p+n\to d$ and  ${\rm {^7Be}}+p\to{\rm{^8B}}$;
   $E_i$  is the relative kinetic energy of the colliding particles;  $\theta$ is the scattering angle  in c.m.s.  and $\lambda=(C_{B;\,1/2}/C_{B:\,3/2})^2$=0.125  \cite{Tab06,Trache2003}. In Eqs. (\ref{eq2}) and (\ref{eq3}),     $b_{B;\,j_B}$  is  the  single particle ANCs for
the  shell model wave function for the   bound   (${\rm {^7Be}}+p$) state,  
 which    determine the amplitude of its tail   \cite{Blokh77};  $b_{d;\,j_d}$ is the amplitude of the tail of the $s$ wave deuteron wave function of     relative motion of the neutron and  proton in the deuteron ($j_d$=1/2);   $\sigma^{\rm{DW}}_{j_B}(E_i,\theta;\,b_{B;\,j_B})$ is the single-particle DWBA cross section.    For independent testing, the single-particle DWBA cross section has also been calculated  using the     LOLA  code \cite{LOLA} restricted only by the first order  perturbation term  of    over    $\Delta V_f^C$ in the transition operator.  Both   calculations gave the same results since  $\Delta V_f^C$=0.  
  In (\ref{eq2}),   the ANC $C_{B;\,j_B}$, the single particle ANCs $ b_{B;\,j_B}$ ($j_B$=1/2 and 3/2)  and the parameter   $b_{d;\,1/2}$    are  unknown, whereas  the values   of   the ANC   for $p+n\to d$  ($C_{d;\,1/2}$)   and  $\lambda$   are  known     \cite{Blokh77,Tab06,MB}. Nevertheless,  the value of the free parameter $b_{d;\,j_d}$ can be determined by  similar way as it was done in  \cite{Azh1999a,Azh1999b,Ar96} as the ``experimental'' ANC value for $p+n\to d$    extracted from the exchange $nd$ and $pd$ scatterings is  known with the high accuracy ($C_{d;\,1/2}^2$=0.774$\pm$0.018 fm$^{-1}$) \cite{Blokh77,MB}. Besides, this value of the  ANC      is in   excellent agreement  with that    calculated  by       different forms of the  realistic   NN potential where its value   was found insensitive to the used form of the NN potential (see \cite{Blokh77} for example).   This circumstance makes it possible  to calculate  the $s$ wave   of   the deuteron wave function  by using    the   Woods-Saxon potential with the geometric parameters chosen   to reproduce  the aforementioned    ANC value of  $C_{d;\,1/2}$ and  to  fix    the value  of $b_{d;\,1/2}$ at 0.942 fm$^{-1/2}$.

To make the dependence of the ${\cal {R}}_{j_B}(E_i,\theta;\,b_{B;\,j_B})$ function on $b_{B;\,j_B}$ more transparent,  we have used    the zero-range version  of DWBA for the  $V_{np}^N$ potential with  fixed optical model parameters in the initial and final states. Note that   this consideration is also valid  for the finite range  of DWBA.  In the radial integral for the matrix element \cite{Austern1966} in the ${\cal {R}}_{j_B}(E_i,\theta;\,b_{B;\,j_B})$ function  we split the space of interaction   of the colliding particles  in  two parts separated by the channel radius $R_{ch}$ \cite{Gon82}: the interior part  (0$\le R\le R_{ch}$), where nuclear forces between the colliding nuclei are important, and the exterior part ($R_{ch}\le R<\infty$), where  the interaction between the colliding nuclei is governed by Coulomb forces only. The exterior part of  the radial integral for the matrix element in the ${\cal {R}}_{j_B}(E_i,\theta;\,b_{B;\,j_B})$ function does not contain explicitly the free parameter $b_{B;\,j_B}$, since for $R>R_{ch}$ the bound [($\rm{^7Be}+p)$] state wave function $\varphi_{B;\,j_B}(R)$[=$\varphi_{B;\,j_B}(R;b_{B;\,j_B})$ \cite{Gon82}] can be approximated by its asymptotic behavior \cite{Blokh77}. Consequently,  parametrization of the differential cross section in the form (\ref{eq2})  allows to fix the contribution from the exterior region in a model independent way, if it is dominant for the peripheral reaction and  if the ANCs $C_{B;\,j_B}$ ($j_B$=1/2 and 3/2) are known. In that case the contribution from the interior part of the radial matrix element to the ${\cal {R}}_{j_B}(E_i,\theta;\,b_{B;\,j_B})$ function, which depends on $b_{B;\,j_B}$ through the fraction $\varphi_{B;\,j_B}(R;b_{B;\,j_B})/b_{B;\,j_B}$ \cite{Gon82,Austern1966},  exactly determines the dependence of  the ${\cal {R}}_{j_B}(E_i,\theta;\,b_{B;\,j_B})$ function on $b_{B;\,j_B}$. It should be noted  that this fraction is convolved with the radial optical wave functions in the initial and final states in the integrand of the  radial  integral for the interior  part of the matrix element. The contribution from the   interior part   to the $d\sigma/d\Omega$ cross section is determined by the free parameters $b_{B;\,j_B}$ and the spectroscopic factors $Z_{B;\,j_B}$  through the product of $Z_{B;\,j_B}^{1/2}\varphi_{B;\,j_B}(R;b_{B;\,j_B})$,  which is really  model dependent due to the unknown free parameters  $b_{B;\,j_B}$.   One notes that the spectroscopic factor ($Z^{1/2}_{B;\,j_B}$), which  is a norm of the radial overlap function  of the bound state ${\rm{^8B}}$ wave function in the (${\rm{^7Be}}$+$p$)-channel,   is related to the  ANC  $C_{B;\,j_B}$    by the equation \cite{Blokh77}
\begin{equation}
\label{eq4}
C_{B;j_B}=Z_{B;\,j_B}^{1/2}b_{B;\,j_B}. 
\end{equation}
  As  is seen from here and will be demonstrated below, at the fixed available  ``indirectly determined''  values of the ANCs $C_{B;j_B}$ (see, Table \ref{table4}),     variation of  single-particle ANC values    $b_{B;\,j_B}$   leads    to the large uncertainty in the absolute values of      the spectroscopic $Z_{B;\,j_B}$. Besides, as a rule, this inaccuracy for $Z_{B;\,j_B}$ can also  grow  because of optical potentials ambiguities  arising mainly in  the interior part of the matrix element.    Apparently, the analogous situation may occur  for other nucleon transfer reactions induced by a deuteron, including   the reactions         systematically   studied in \cite{Liu2004}  taking into account      the deuteron's break up  in the field of the target. In this connection,  it should  be noted  that   for small relative  distances between  colliding light nuclei, which are responsible for the low partial-wave  amplitudes   corresponding to processes proceeding inside nuclei,  the optical model potentials cannot, generally speaking, reflect the true nature of many-particle nuclear interactions \cite{Mukh97}.  Therefore,    the dependence on the single-particle ANCs and optical model potentials in  the interior part of the matrix element can be one of the main reasons of the strong dependence of empirical (``experimental'') values of the spectroscopic factors $Z_{B;\,j_B}$ extracted from  different forms of   the    DWBA analysis (see,  Refs. \cite{Ogata2006,Ogata2003}).

 Nevertheless, if the ${\rm{^7Be}}$($d$,$n$)${\rm{^8B}}$ reaction  is peripheral in the angular region  near the main peak the contribution of the internal part into the ${\cal {R}}_{j_B}(E_i,\theta;\,b_{B;\,j_B})$ must strongly be suppressed. In this case, Eq. (\ref{eq2}) can be used for determination of the square ANCs  $C^2_{B;\,j_B}$,  since, in the external part of the matrix element, the  optical potential ambiguity and the dependence of  the ${\cal {R}}_{j_B}(E_i,\theta;\,b_{B;\,j_B})$ function on $b_{B;\,j_B}$ can be reduced to minimum. To this end,   according to  \cite{Gul1995},   at the    fixed values
of  $ b_{d;\,1/2}$, $\lambda$ and optical model parameters for the initial and final states, obviously the peripheral character for the ${\rm {^7Be}}$($d$,$n$)${\rm {^8B}}$ reaction
   is  conditioned by
\begin{equation}
\label{eq5}
{\cal {R}}_{j_B}(E_i,\theta;\,b_{B;\,j_B})= f(E_i,\theta),
\end{equation}
   where the  left-hand side of Eq. (\ref{eq5}) should not depend on $b_{B;\,j_B}$  for each fixed   energy $E_i$   and   scattering angle $\theta$ belonging to   the main peak region.  Then from (\ref{eq2}) and (\ref{eq5}) the following condition
\begin{equation}
\label{eq6}
C^2_{B;3/2}=\frac{d\sigma/d\Omega}{{\cal {R}}_{3/2}(E_i,\theta;\,b_{B;3/2})+\lambda{\cal {R}}_{1/2}(E_i,\theta;\,b_{B;1/2})}=const
\end{equation}
 must be fulfilled  for each fixed energy $E_i$, $\theta$   and  the function of $ {\cal {R}}_{j_B}(E_i,\theta; \,b_{B;\,l_Bj_B})$ from (\ref{eq5}).
 
Thus, introduction of  the conditions
(\ref{eq5}) and  (\ref{eq6})  into the DWBA  analysis  guarantees the correct absolute normalization of the  peripheral reaction  cross section  and      supports   the assumption about   the dominance    of the peripheral character of the proton transfer        within (or near) the main peak region  of the angular distribution,  which  is mainly  determined by   the true  peripheral partial-wave amplitudes at  $l>>$1  \cite{Igam072}. Therefore,    fulfilment (or weak violation   within the  errors of   the  experimental  differential cross section  $d\sigma^{exp} /d\Omega$) of the conditions (\ref{eq5}) and  (\ref{eq6})  makes it possible    to obtain the experimental (``indirectly determined'') value of squared ANC $(C_{B\,j_B}^{exp})^2$    for ${\rm{^7Be}}+p\to{\rm{^8B}}$  using  the  $d\sigma^{exp} /d\Omega$,   measured in the  main peak of the angular distribution,   for $d\sigma /d\Omega$,   the values of $C^2_{d;\,1/2}$ \cite{Blokh77,MB} and the parameter $b_{d;\,1/2}$ given above in the right hand side (r.h.s.) of (\ref{eq6}).

\begin{center}
{\bf B. Asymptotic normalization coefficients for   ${\rm {^7Be}}+p\to{\rm{^8B}}$}
\end{center}

\hspace{0.6cm}   To determine the ANC values for ${\rm{^7Be}}+p\to{\rm{^8B}}$, the experimental differential cross sections   measured in   inverse kinematics    for the ${\rm{^7Be}}(d,n){\rm{^8B}}$  reaction   at the energy  of   $E_i$=4.5  MeV \cite{Das2006}  were reanalyzed  by the modified DWBA.

\begin{figure}
\begin{center}
 \includegraphics{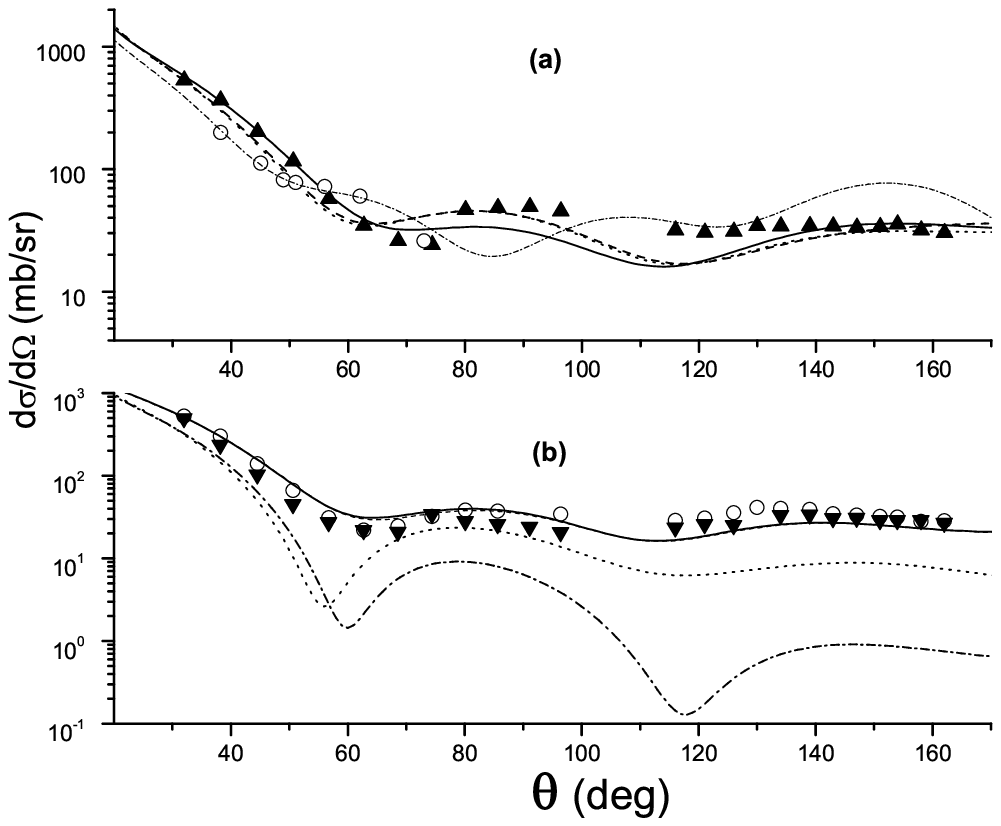}
\end{center}
 \caption{\label{fig1} The fit of the elastic $d+{\rm{^7Li}}$ and  $d+{\rm{^7Be}}$ scattering cross sections  by using  different sets of  optical potentials.   In ($a$), the solid,  dashed and  dotted lines correspond to the sets 4, 3 and 2, whereas the dash-dotted line corresponds to the set S1 from \cite{Das2006}; experimental data   for  the elastic $d+{\rm{^7Li}}$ scattering    at $E_d$=6 MeV (closed  triangles) and for  the   $d+{\rm{^7Be}}$ scattering $E_d$=5.79 MeV (open circles) are taken from  \cite{Avrig2005} and \cite{Das2006}, respectively. In ($b$), the solid and dashed   lines correspond to the sets 1$a$ and 2$a$ of the present work, and   the dash-dotted and dotted    lines correspond to the sets 1 and 2  recommended in  \cite{WL1998};   experimental data (open and close points) for  the elastic $d+{\rm{^7Li}}$ scattering  \cite{Avrig2005}    at $E_d$=7.0 and 8.0 MeV, respectively.}
 \end{figure}

  Four      sets of the optical potentials listed in Table \ref{table1} were used.  These were obtained from the global parametrization given in [35--39] and a $\chi^2$ minimization analysis   by means of the best fits to the experimental $d+{\rm{^7Li}}$, $n+{\rm{^9Be}}$  and $n+{\rm{^{11}B}}$ scattering angular distributions  in the forward hemisphere     at the corresponding projectile  energies. Such a way of the fitting provides equally    good reproduction of  the experimental angular distribution within the main peak region   as it will be shown below.   As an illustration, for     the  sets 2--4 of the optical potentials,  Fig. \ref{fig1}$a$   shows  the results of comparison between  the calculated  angular distributions   and    the experimental     data   for elastic   $d+{\rm{^7Li}}$  scattering taken from   Ref.\cite{Avrig2005} (closed triangles)    at most near   corresponding   energies. As is seen from this figure,       the used sets reproduce well the corresponding  experimental  angular distributions up   to $\sim$90$^o$.  A similar result is obtained for  set 1 of Table \ref{table1}. Besides,   in Fig  \ref{fig1}$a$,   the result of calculation (dash-dotted line), obtained   with  the optical potentials for the set $S1$ recommended in Ref. \cite{Das2006}, and its comparison with the experimental     data \cite{Das2006} (open circle)      are  presented for the elastic $d+{\rm{^7Be}}$ scattering  at $E_i$=4.5 MeV.  One can see that  the  optical potentials recommended in  \cite{Das2006}   describe well  the experimental    data  only  in    the    narrow  angle range of the forward hemisphere. In this connection,  we note that the deuteron groups corresponding to the ground and  first excited states were not resolved in the experiment.  Moreover, a    comparison of behavior of the experimental angular distributions of the elastic deuteron scattering  on the nuclei   ${\rm{^7Be}}$ and ${\rm{^7Li}}$ shows the  overestimation    of values of the elastic cross section  in the angular range 55--75$^o$ in \cite{Das2006}. It may be the result of the underestimation of  the cross section  of the inelastic scattering with the formation of   the 427 keV excitation level  in ${\rm{^7Be}}$ in the angle range mentioned above.  Indeed, as  can be seen from Fig. 5 of Ref. \cite{Avrig2005}, at the elastic deuteron scattering on the nucleus ${\rm{^7Li}}$, the first minimum of the angular distribution is smoothly displaced towards small angles  increasing the relative energy (approximately from 80$^o$ at 4 MeV to 70$^o$ at 5 MeV). The value of the differential cross section for the ${\rm{^7Be}}(d,d_o){\rm{^7Be}}$ data at 4.5 MeV \cite{Das2006}  changes by  $\sim$2.4 times between the scattering angles 60--75$^o$ (i.e., on the left slope of the expected first minimum of the angular distribution), still   not reaching the minimum. Whereas at the same energy (from the data averaged on energy at 4 and 5 MeV \cite{Avrig2005}), the differential cross section for ${\rm{^7Li}}(d,d_o){\rm{^7Li}}$ scattering changes only by $\sim$  1.3 times, reaching the first  minimum at an angle  $\sim{\rm {75}}^o$.   It should be noticed that the position of the first minimum depends very smoothly on the mass number at the fixed   energy (see for example, work \cite{Fitz1967}), therefore the trend of the angular distribution and position of the first minimum for deuteron scattering on  ${\rm{^7Li}}$ and ${\rm{^7Be}}$ nuclei should be similar. For specification of the cross section behavior in the region of the first minimum of angular distribution, which is an essential criterion for the realistic OMP selection, the experiment with a smaller energy dispersion of the ${\rm{^7Be}}$  beam is needed. Taking into account the above-stated considerations, we do not include   the OMPs recommended in  \cite{Das2006} for the analysis of the ${\rm{^7Be}}$($d$, $n$)${\rm{^8B}}$  reaction performed below.

For  calculation of the shell model two-body bound ${\rm{^8B}}$(${\rm{^7Be}}+p$) state wave functions the  Woods-Saxon potential is   used  with varying the geometric parameters within  the physically acceptable limit  by means of adjusting the well  depth  to  the experimental binding energy (137 keV) for each of  the geometric parameters. At first,  we have tested validity  of the condition (\ref{eq5})  for the first six   experimental  points    of  the angular distribution of the      reaction  presented in  \cite{Das2006}  and   for all   sets of the   optical potentials of Table \ref{table1}. This test is done by changing the geometric parameters $r_o$ and $a$  of the  Woods-Saxon potential, used for calculation of the  bound (${\rm{^7Be}}+p$) state wave function, in a wide physically acceptable ranges ($r_o$=1.10--1.40 fm and $a$=0.50--0.75 fm) with respect to their ``standard'' values ($r_o$=1.25 fm and $a$=0.65 fm). Such   variation of the $r_o$ and $a$   results in changing  the single-particle ANCs ($b_{{\rm{B}};\,j_B}=b_{{\rm{B}};\,j_B}(r_o,\,a)$ \cite{Gon82}) with  $j_B$=1/2 and 3/2) within the intervals  of  0.615$\le b_{{\rm{B}};\,3/2}\le$0.795 fm$^{-1/2}$ and  0.596$\le b_{{\rm{B}};\,1/2}\le$0.782 fm$^{-1/2}$. The calculations show that  over these   intervals   the  condition (\ref{eq5})  for each angle $\theta$ within  the main peak region is fulfilled within  $\pm$1.5\%, whereas  $\sigma^{\rm{DW}}_{l_dl_B}(E_i,\theta;\,b_{{\rm{B}};j_B})$ entering (\ref{eq3}) is a rapidly varying function  of $b_{{\rm{B}};\,j_B}$ for $j_B$=1/2 and 3/2.

\begin{figure}
\begin{center}
 \includegraphics{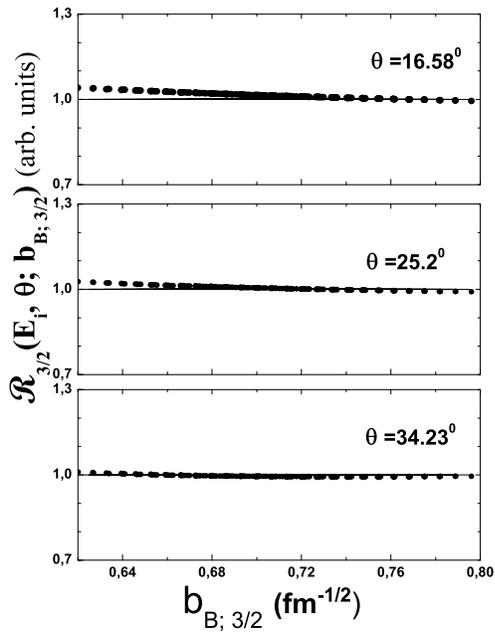}
\end{center}
 \caption{\label{fig2} The dependence of ${\cal {R}}_{3/2}(E_i,\theta;\,b_{B;\,3/2})$ on  the single-particle $b_{B;\,3/2}$   at the different    angles  $\theta$ for  the   energy of $E_i$=4.5 MeV  and with potential parameters of the set 3 in Table \ref{table1}.   The width of the bands for fixed values of  $b_{B;\,3/2}$ corresponds to     variation of the   parameters $r_o$ and $a$ of the adopted Woods-Saxon potential within the intervals from $r_o$=1.10 to 1.40 fm and $a$=0.50 to 0.75 fm.}
\end{figure}

Fig. \ref{fig2} shows a plot of the ${\cal {R}}_{3/2}(E_i,\theta;\,b_{B;\,3/2})$ dependence on the single-particle $b_{{\rm{B}};\,3/2}$ for the potential of the set 3 in Table \ref{table1}  within the aforementioned interval for three     angles   $\theta$ within  the main peak.    The width of the band for these curves is the result of the weak ``residual'' ($r_o,a$) dependence of  ${\cal {R}}_{3/2}(E_i,\theta;\,b_{B;\,3/2})$  on the parameters  $r_o$ and $a$ (up to $\pm$1\%) for $b_{{\rm{B}};\,3/2}(r_o,a)$=const \cite{Gon82}. The same dependence is also observed for $j_B$=1/2. For example,      the arithmetic averaged values of the ${\cal {R}}_{3/2}(E_i,\theta;\,b_{B;\,3/2})$ and ${\cal {R}}_{1/2}(E_i,\theta;\,b_{B;\,1/2})$ obtained   in the     intervals  for    $b_{B;\,j_B}$ ($j_B$=1/2 and 3/2) mentioned above are equal to  49.76$\pm$0.04 and 72.31$\pm$0.18 mb$\cdot$fm/sr  at   $\theta$=16.58$^o$  and equal    to 25.61$\pm$0.01 and 39.41$\pm$0.12 mb$\cdot$fm/sr, respectively, at $\theta$=29.6$^o$. Here,  the pointed out the uncertainties are the averaged square errors (a.s.e.), which  involve those arising due to the observed weak dependence of these calculated functions  with  changing of $b_{p{\rm{^7Be}};\,j_B}$ and their   weak ``residual'' ($r_o,a$) dependence of ${\cal {R}}_{3/2}(E_i,\theta;\,b_{B;\,j_B})$ at $b_{B;\,j_B}$=const.
It is seen that, at   fixed values of  $\theta$ and $E_i$ the averaged  values of the ${\cal {R}}_{j}(E_i,\theta;\,b_{B;\,j_B})$  functions do  not depend practically  on the free parameter $b_{B;\,j_B}$. The same situation  is observed for the other considered  values of $\theta$ and  the sets of the optical potentials from Table \ref{table1}.

\begin{figure}
\begin{center}
\includegraphics{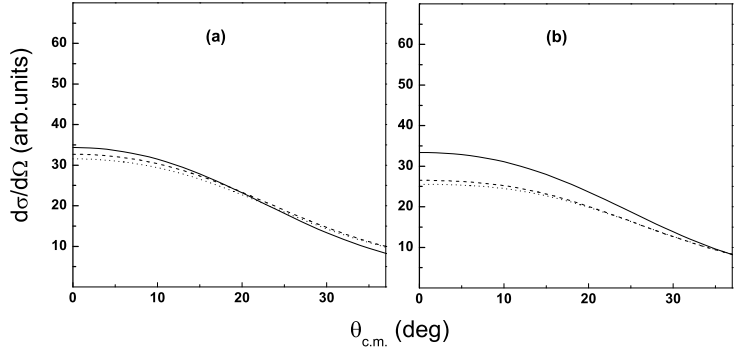}
 \end{center}
 \caption{\label{fig3}
 Angular distributions at different  cutoff radii
($R_{cut}$)   for $E_i$=4.5  MeV (a) and $E_i$=5.8  MeV (b) calculated in DWBA with
potential parameters of the sets 3 and  1$a$ (Table \ref{table1}), respectively.
The solid, dotted and dashed lines are the differential  cross
sections calculated for the cutoff radius $R_{cut}$=0.0, 4.0 and 5.0
fm, respectively.}
\end{figure}

\begin{figure}
\begin{center}
\includegraphics{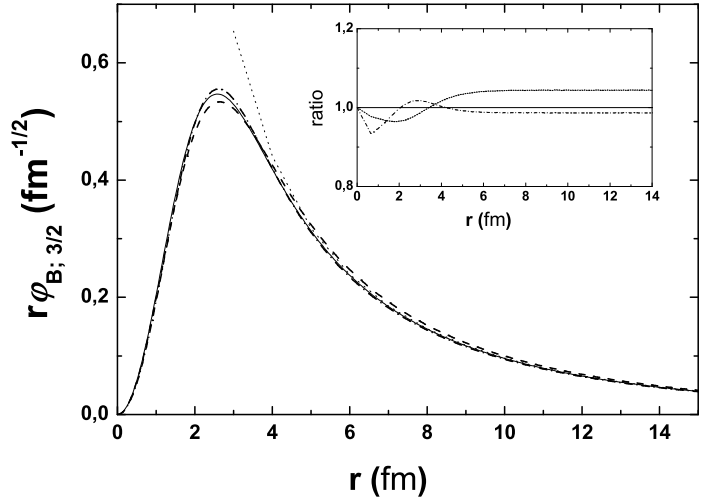}
\end{center}
 \caption{\label{fig4} The radial behaviour of the single-particle    ${\rm{^8B}}$[=(${\rm{^7Be}}+p$)] bound  state wave function  $r\varphi_{B;\,j_B}(r)$ with  $j_B$=3/2  calculated for the Woods-Saxon
 potential with different sets of ($r_o,\,a$)-pair and $b_{B,\,3/2}$:(1.00 fm; 0.50 fm) and 0.6200 fm$^{-1/2}$ (the dashed line),(1.25 fm; 0.65 fm) and 0.7679 fm$^{-1/2}$ (the solid line), (1.40 fm; 0.80 fm) and 0.7960 fm$^{-1/2}$ (the dashed-dotted line). The Coulomb radius $r_C$=1.30 fm. The dotted  line is the tail
  $b_{B,\,3/2}W_{-\eta_B;3/2}(2\kappa r)$
  of the bound state wave function $r\varphi_{l_Bj_B}(r)$
     with $b_{B,\,3/2}$=0.7679 fm$^{-1/2}$($r_o$=1.25 fm and $a$=0.65 fm).}
\end{figure}

We also performed   calculations of the DWBA cross sections (\ref{eq2}) at
forward angles  for different values of the cutoff radius $R_{cut}$
(lower limit in radial integration over the distance ($R$)  between centers of masses of 
the colliding particles) to check in an independent   manner  the peripheral character of the
${\rm{^7Be}}(d,n){\rm{^8B}}$  reaction at energies $E_i$=4.5    MeV.
  Calculations have been done  for all
sets of the optical potentials of Table \ref{table1} and   the
Woods-Saxon potential  for the bound state of  ${\rm{^8B}}$ with the standard  geometric
parameters ($r_o$=1.25 fm and   $a$=0.65 fm)  and   the Thomas
spin-orbital term. The dependence of the DWBA cross
sections  on the $R_{cut}$, $\frac{d\sigma}{d\Omega}(R_{cut})$  is
shown in Fig. \ref{fig3}$a$ for   set 3.    It shows that the
contribution  of the region with $R\le R_{cut}\lesssim$ 5.0 fm into
the calculated DWBA  cross section is strongly suppressed and
practically negligible.  The bound
state wave functions $r\varphi_{B;\,j_B}(r)$  ($j_B$=3/2) of the ${\rm{^8B}}$
nucleus in the (${\rm{^7Be}}+p$)-channel calculated for   different values of the
geometric   parameters  reach their asymptotic behaviour
$b_{B,\,3/2}W_{-\eta_B;3/2}(2\kappa r)$ for $r\gtrsim$ 5.0 fm, where
$W_{-\eta_B;3/2}(x)$ is the Whitteker function; $\eta_B$ is the
Coulomb parameter  for ${\rm{^8B}}$=(${\rm{^7Be}}+p$) bound state
and $\kappa=\sqrt{2\mu_{p{\rm{^7Be}}}\varepsilon}$ in which
$\varepsilon$ is the binding energy of  ${\rm{^8B}}$
in  (${\rm{^7Be}}+p$)-channel. Fig.  \ref{fig4} demonstrates the
dependence of the bound   state wave function
$r\varphi_{B;\,j_B}(r)$  ($j_B$=3/2)  
  on the geometric parameters
$r_o$ and $a$ of  the Woods-Saxon potential.

Testing  the condition (\ref{eq5}) and calculating  the DWBA
cross sections (\ref{eq2})  at the first four experimental points of the angular distribution
also were done  for different values of the cutoff radius $R_{cut}$    for
the        ${\rm{^7Be}}(d,n){\rm{^8B}}$  reaction at the  energy
$E_i$=5.8  MeV \cite{WL1998}.  The calculations were performed for the  sets
1$a$, 2$a$ and 3$a$ of the optical potentials obtained by means of
 fitting experimental scattering data  closest to the kinematics of \cite{WL1998} energies in the same way as it is done above.
  The results  are   presented  in Table \ref{table1}.  Fig. \ref{fig1}$b$ shows  the results of comparison  of   the angular distributions calculated  for the sets  1$a$ and 2$a$ (solid and dashed lines, respectively) with    the experimental     data   for elastic
$d+{\rm{^7Li}}$  scattering taken from Ref.\cite{Avrig2005} (open
circles and closed triangles for  $E_d$=7.0 and 8.0 MeV,
respectively). As it is seen from the figure,        the used sets
reproduce well the corresponding experimental  angular distributions
up   to $\sim$90$^o$, whereas the optical potentials recommended in
\cite{WL1998} provide poor fits to the experimental data (see
dash-dotted and dotted    lines in Fig. \ref{fig1}$b$). The same
situation  is observed  for  set  3$a$. But, the
calculations performed for all the considered sets of the optical potentials  show that
the condition (\ref{eq5}) is fulfilled within $\pm$15.0\% over the mentioned above
 intervals  for $b_{{\rm{B}};\,3/2}$   and $b_{{\rm{B}};\,1/2}$
for each angle $\theta$ within  the main peak region.  
The same situation was observed in  \cite{Toj2014} where  the
optical potentials of \cite{WL1998} were  used. The dependence of
the DWBA cross sections  on  $R_{cut}$ at the forward angles  
is shown in Fig. \ref{fig3}$b$ for   set 1$a$. As  is seen from the
figure,   the contribution  of the region with $R\le
R_{cut}\lesssim$ 5.0 fm  to the calculated DWBA  cross section is
noticeable    and occurs to be  about of 30\%. The same result is obtained for the
sets 2$a$ and 3$a$.

  It follows from here that the
${\rm{^7Be}}(d,n){\rm{^8B}}$  reaction at   the  energy   $E_i$=5.8
MeV is not  purely   peripheral due to the fact that the contribution of the interior part of the matrix element   into the ${\cal {R}}_{3/2}(E_i,\theta;\,b_{B;\,j_B})$ function is significant. One notes once more that    the  strong  dependence of this function  on $b_{B;\,j_B}$ is mainly associated    with the interior part of the matrix element   and is determined     by  the bound state  $\varphi_{B;\,j_B}(r;\,b_{B;\,j_B})$ wave function, which is  displayed in  Fig. \ref{fig4} for $j_B$=3/2 and  three fixed values of $b_{B;\,j_B}$ ($b_{B;\,j_B}$=0.620; 0.768  and 0.796 fm$^{-1/2}$).   For these values of $b_{B;\,j_B}$ the calculated wave functions change noticeably in the interior region, whereas the observed discrepancy grows from 3\% to 8\% as the relative distance between the ``valence'' proton  and the center mass of the core (${\rm^7Be}$) decreases from 3.0 fm to 0.2 fm (see, the insert in Fig. \ref{fig4}). 

Therefore, here in reality  one should take into account  the effect of a node at short      distance in the bound state wave function  which is due to Pauli antisymmetrization.  Besides, as  is mentioned above,   the additional difficulty, connected   with     uncertainties  of  the  optical model potentials in  the interior part of the matrix element for description of  the elastic scattering of lightest  particles on light nuclei  \cite{Gon82,Liu2004,SMSZ1967},  can be faced.   The facts mentioned above can apparently be one of  the   reasons why the value of  the spectroscopic factor for  ${\rm{^8B}}$  in the (${\rm{^7Be}}+p$) configuration, extracted in \cite{Ogata2003} from the ${\rm{^7Be}}(d,n){\rm{^8B}}$  analysis at   the  energy   $E_i$=5.8 MeV with  using different sets of the input data, has a large spread (about 30\%).   
   Therefore, the analysis of  the experimental data of  \cite{WL1998} performed within the modified
DWBA in \cite{Toj2014} and in the present work  as well as that performed in \cite{Ogata2003}  cannot allow one to    obtain the reliable ANC values for   astrophysical application. In contrast to this case, as it is
demonstrated above, the ${\rm{^7Be}}(d,n){\rm{^8B}}$  reaction at
the  energy   $E_i$=4.5  MeV is  predominantly    peripheral in the main peak region of the angular distribution.    At this case,  the influence of  the effects connected with  the true structure of many-particle wave functions of the   nuclei (deuteron, ${\rm{^7Be}}$ and ${\rm{^8B}}$),  exhibited     mainly in the interior part of the matrix element, can be ignored. So, in the surface and   outer regions of the nucleus,  the  wave functions of the relative motion at   the initial   and final   states can be described by the optical model, and the bound [(${\rm{^7Be}}+p$)] state wave function     by the one-particle shell model wave function    in a correct manner reducing their uncertainties to a minimum.     
 Consequently,  the experimental data of  \cite{Das2006} are used  for
obtaining   the ``indirectly determined'' values of the ANCs for ${\rm{^7Be}}+p\to{\rm{^8B}}$.

\begin{figure}
\begin{center}
\includegraphics{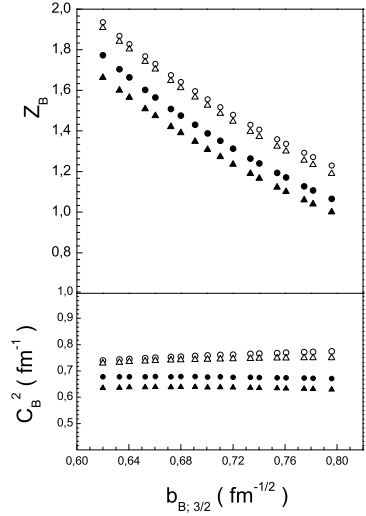}
 \end{center}
 \caption{\label{fig5}
Values of the spectroscopic factors   $Z_B$ (top)
  and the squared ANC $C^2_B$ (bottom)  as a function of the single-particle $b_{B;\,3/2}$ obtained from    the modified DWBA  ${\rm{^7Be}}(d,n){\rm{^8B}}$ analysis  at the   energy    $E_i$=4.5    MeV for the different fixed    angles  $\theta$ and the sets of the optical potentials.   Data denoted by   $\circ $  and $\vartriangle$ ($\bullet$ and $\blacktriangle$)  correspond to  the experimental points $\theta$ of 16.6$^o$ and 29.6$^o$, respectively, for set 3(2).  }
\end{figure}

The condition  (\ref{eq6}) was used for  each    $\theta$ from    the main peak region  of the angular distribution as only in such a case is  the absolute value of  the differential cross section defined by  the ``experimental''   value of the  ANC. For  illustration,
 we present in Fig. \ref{fig5} the results of a calculation of  the ratio  in the r.h.s. of the expression (\ref{eq6}), where instead of the calculated differential cross section  the corresponding  experimental cross sections    are taken. The calculations were performed  for $\theta$=16.6$^o$ and 29.6$^o$  and the sets 2 and 3 of the optical potentials.
Here, it was taken into account  the fact   that the ratio $\tilde{R}=b_{B;\,1/2}(r_o,\,a)/b_{B;\,3/2}(r_o,\,a)$ does not depend practically  on    variation of  the free $r_o$ and  $a$ parameters of  the  Woods--Saxon potential   ($r_o$ ranging from 1.10--1.40 fm and $a$  in the range of 0.50--0.75 fm).  The  value of $\tilde{R}$ is equal to 0.9742$\pm$0.0065.   It is seen from the figure that the $C^2_B$ values     are   weakly dependent on the    $b_{B;\,3/2}$  value.  However, the values of the spectroscopic factor $Z_B$ determined by the relation $Z_B$=1.006$C^2_B/b_{B;\,3/2}^2$, which can be obtained from (\ref{eq4}) and using the aforementioned values for $\tilde{R}$ and   $\lambda$, change strongly (Fig. \ref{fig5} (top)).  We found that    the same dependences for $C^2_B$ and $Z_B$   occur  
 for all   other considered  scattering angles   and   the sets of the optical potentials from Table \ref{table1}.

\begin{figure}
\begin{center}
\includegraphics{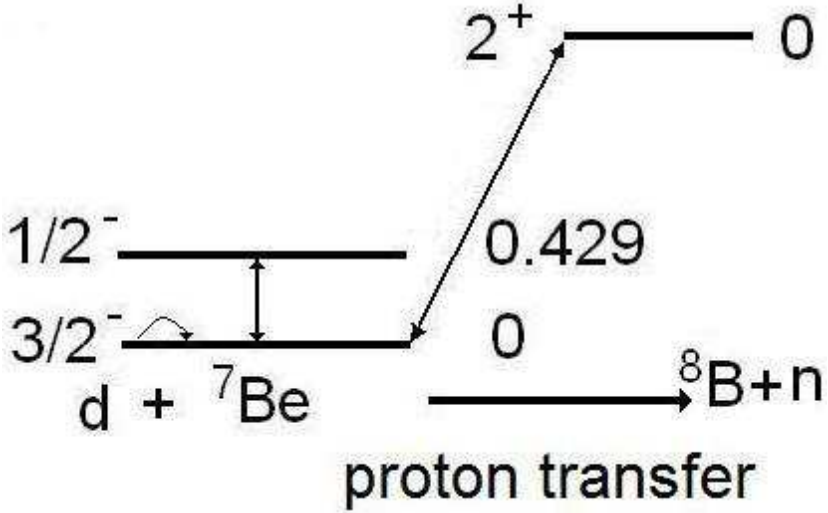}
  \end{center}
 \caption{\label{fig6} The coupling scheme used in calculations of the cross section for the  ${\rm{^7Be}}(d,n){\rm{^7Be}}$ reaction by the coupled reaction channels  method.}
 \end{figure}

 To increase the  accuracy of the ANC values required for their astrophysical application,     the contribution of    the compound (${{\rm^9B^*}}$)  nucleus (CN) and coupled channel effects (CCE) connected with the   elastic and inelastic $d+{\rm{^7Be}}$ scattering should be taken into account,   similar as it was done in Refs. \cite{WL1998,Das2006} and \cite{Nunes2001,Burt2013}.  The influence of the CN   contribution $\Delta_{CN}$    on  the ANC values extracted     for  each    $\theta$ from the main peak region   has been determined using the results of work \cite{Das2006}, where $\Delta_{CN}=2|d\sigma_{CN}/d\Omega-d\sigma/d\Omega|/(d\sigma_{CN}d\Omega+d\sigma/d\Omega)$
in which $d\sigma_{CN}/d\Omega$ is the  CN cross section.   The    contribution of $\Delta_{CN}$ 
increases with  increase of $\theta$ in  the main peak region what leads to a decrease of the ANC values, obtained from the relation (\ref{eq6}). For example, it is   by 0.8\% at $\theta$=8.2$^o$ and by 2.9\% at $\theta$=34.2$^o$ for the set 1 and by 1.0\% at $\theta$=8.2$^o$ and by 3.1\% at $\theta$=34.2$^o$ for the set 3. The same results practically occur for other sets of the used optical potentials.  As a whole the CN   contribution to the  cross sections \cite{Das2006} in the main peak region   is quite small and, consequently, it results in barely changing  the extracted ANC values.  The CCE contributions   to the DWBA cross sections    for each experimental point of $\theta$ from the main peak region,  denoted  by $\Delta_{CCE}$=2$|d\sigma_{CCE}/d\Omega-d\sigma/d\Omega|/(d\sigma_{CCE}/d\Omega+d\sigma/d\Omega)$ \cite{Nunes2001}, where   $d\sigma_{CCE}/d\Omega$ is the coupling channel cross section,   has been determined using the FRESCO computational code \cite{Thom2006}.   The coupling  between the ground ($E^*$=0.0; $J^{\pi}$=3/2$^{-}$) and first excited ($E^*$=0.429 MeV;$J^{\pi}$=1/2$^{-}$) states of the ${\rm{^7Be}}$ nucleus was calculated  with the collective form factor of the rotational model for the quadrupole  transition, as it was done in \cite{Burt2013} (see the coupling scheme  in Fig. \ref{fig6}). The spectroscopic amplitudes $A_j$     are taken from Ref. \cite{WL1998} ($A_j=Z_j^{1/2}$).   For the core-core interaction  the optical potentials adopted for   the exit channel of the analyzed reaction are used since  there are no optical potentials for the $n{\rm{^7Be}}$-scattering.  The deformation length $\delta_2$  is   taken equal to   2.0 fm, which results in a change  of the deformation parameter $\delta_2$(=$ \beta_2 R_{Be}$, where $R_{Be}=r_o7^{1/3}$)  from 1.02 to 1.04 in a dependence on the parameter radius $r_o$ for  the real parts of the used optical potentials.  The averaged value of the $\beta_2$ is   equal to 1.03, which is in a good agreement with the value of 1.00  obtained in \cite{Burteb1996} from the analysis of   $\alpha+{\rm{^7Li}}$-scattering. The   results are shown in Table \ref{table2} where the  coupling channel differential cross sections and the coupled channel   corrections (percentagewise) are given in  the forward hemisphere and all   sets of the optical potentials used are shown. As it is seen from Table \ref{table2}, for all the used sets of the optical parameters the influence of    contribution of  the CCE on the ANC values  becomes larger as the scattering angle increases.  For example, taking into account  the
CCE   contribution   increases  the ANC values   by 2.1\% at $\theta$=8.2$^o$ and by  9.4\% at $\theta$=34.2$^o$ for the set 1 and decreases  the ANC values by 1.0\% at $\theta$=8.2$^o$ and by 9.0\% at $\theta$=34.2$^o$ for the set 3. Thus, the detailed study of the peripheral character of   the considered reaction       makes  it possible   to extract the values of  ANCs  $C^2_{B;\,3/2}$ and  $C^2_{{\rm B}} $[=$C^2_{B;\,3/2}$(1+$\lambda$)]  by using in the r.h.s.  of the relation (\ref{eq6}) the experimental differential cross sections  and values of the function  ${\cal{R}}_{j_B}$ for the different     scattering angles from  the forward hemisphere. At this,    the contribution of  the compound nucleus and  the CCE to the extracted ANC values  should be   taken into account.

\begin{figure}
\begin{center}
\includegraphics{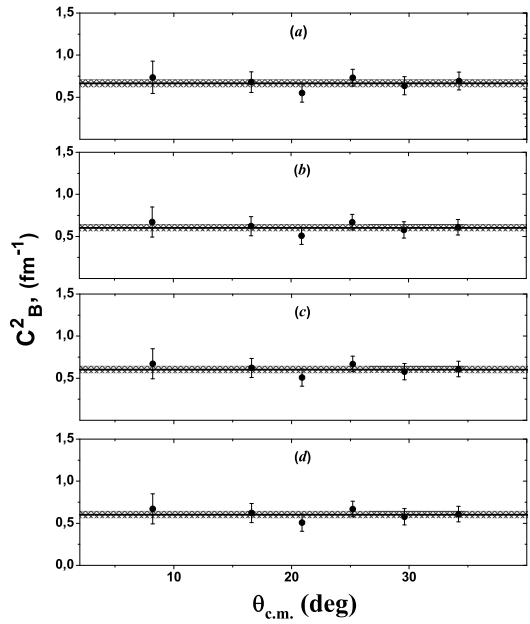}
\end{center}
 \caption{\label{fig7} The values of the squared ANC $C^2_B$ for ${\rm{^7Be}}+p\to{\rm{^8B}}$  for each of the experimental   $\theta$.  Data in (a)--(d) are obtained from the analysis of the experimental differential cross sections  of Refs. \cite {Das2006} at $E_i$=4.5 MeV     and   the  sets 1--4 of the optical potentials, respectively. The solid lines present the results for the weighed mean values. The widths  of  the bands are the corresponding weighed uncertainties.}
\end{figure}

 The results of the ANC obtained for each       experimental  $\theta$ point   and corresponding to the sets 1--4 are presented in Fig. \ref{fig7} ($a$)--($d$), respectively. The uncertainties pointed out in this figure correspond to those found from Eq. (\ref{eq6}) (a.s.e.), which includes both the experimental errors in the corresponding $d\sigma^{exp}/d\Omega$  and mentioned uncertainties in the  ${\cal {R}}_{j}(E_i,\theta;\,b_{B;\,j})$  functions. The solid line and the width of the band present the results for the weighted mean values and their uncertainties \cite{Angolo1999}, respectively.   It is seen from  Fig. \ref{fig6} that the r.h.s. of the relation  (\ref{eq6}) practically does not depend on the angle $\theta$ although  absolute values of the experimental cross sections   depend noticeably on the angle $\theta$ and change by up to the factor 2.6 with    $\theta$  changing from 8.2$^o$ to 34.2$^o$.

The weighted mean values  of the  squared ANCs obtained from all  ANC values corresponding   to different values of $\theta$ in the forward hemisphere  are presented   in  the lines  1--8 of    Table \ref{table3}  for the sets 1--4 of the optical potentials. The uncertainties shown  for each   set of the optical potentials correspond to the weighted uncertainty, which includes only the experimental errors in the measured  differential cross sections.  There,  the   weighted mean values   of the squared ANCs extracted without taking into account the CN  and CCE contributions are also presented in parenthetical figures. An influence of the CN  and   CCE contributions on the ANC values is also observed.  As is seen from Table \ref{table3}, the weighted mean value of the ANCs   depend     weakly on    the set of the optical parameters used, and their spread is within the experimental uncertainties for  the $d\sigma^{\rm{exp}}/d\Omega$ \cite{Das2006}.  The averaged    value of the squared ANCs   $C^2_B $ recommended in this work is presented     in the ninth and eleventh lines of  Table \ref{table3} in which the theoretical uncertainty is  the a.s.e. connected with the used sets   of  the optical potentials  and the   uncertainty  in the $ {\cal {R}}_{j_B}(E_i,\theta;\,b_{B;\,j_B})$ mentioned above. There the experimental uncertainty is the averaged arithmetic error obtained from the errors for each   set of the optical potentials. The overall uncertainties, which are the a.s.e. of the experimental and theoretical ones, are presented in the eleventh  and twelfth lines. It is seen from Table \ref{table3}, the CN and CCE contributions change  the ANC values about  4\%   and the overall uncertainty for the derived ANCs is about 7\%, whereas the theoretical and experimental ones are about  3.5\% and  6.1\%, respectively, i.e., the experimental uncertainty is fairly larger (a factor  about 1.7) than the theoretical one. The ANC value recommended in the present work
 is    $C^2_B$=0.626$\pm$0.022(th)$\pm$0.038(exp) fm$^{-1}$ or $C^2_B$=0.626$\pm$0.044 fm$^{-1}$ with  the overall uncertainty, which    are also listed in the second and third  lines of  Table  \ref{table4} together with those obtained by the other authors in [10,16,20,24,46--50]. As  is seen from Table \ref{table4}, the weighted mean value of  $C^2_B $, obtained in the present work,    is in    excellent  agreement   with that of \cite{Igam08}, which was  derived    from the independent analysis of the experimental ${\rm{^7Be}}(p,\gamma){\rm{^8B}}$ astrophysical $S$ factors at extremely low energies.  However, the  obtained value for $C^2_B$        differs noticeably from that recommended in \cite{Tab06,Ogata2003,Huang2010,Bar95}.  As we discussed above,    the ANC value of \cite{Tab06} is    model dependent \cite{Yarm2013,Igam072, Ogata2006,Baby2003} on account of  using  the first order over the $\Delta V_f^C$ potential.   The ANC value of \cite{Ogata2003} and the discussed  theoretical uncertainty of  it, which can easily be obtained     using the   $S_{17}(0)$ value derived in \cite{Ogata2003} (see, Table \ref{table4}) and its  relation with the ANC given  there,   is also model dependent  because of    the fairly  large  ambiguity    for the spectroscopic factors.  The results of \cite{Huang2010}  and \cite{Bar95}  were obtained  from the ${\rm{^7Be}}(p,\gamma){\rm{^8B}}$ $R$-matrix analysis, where   the direct part of the amplitude is expressed in terms of the channel reduced width and is also determined by a model dependent way.  In reality, the ANC value obtained in \cite{Huang2010}  and \cite{Bar95} can have the uncertainty arising due  to an ambiguity in values determined by fitting all the other free parameters used. 
 In addition, as  is  seen from Table \ref{table4}, the ANCs obtained by us   also differ noticeably from the values  of $C^2_B $ derived in \cite{Ogata2006,Trache01} from  the  ${\rm {^{208}Pb(^8B,p^7Be)^{208}Pb}}$ breakup reaction.     In \cite{Ogata2006},   this process is analysed, as noted above, within the strict three-body model based on the CDCCM in which  all nuclear and Coulomb interactions are exactly taken into account   in  the all-order perturbation  theory in the transition operator. Nevertheless, firstly, the cross section of the considered  breakup process is expressed in terms of the spectroscopic factor $Z_B$ (denoted by $\alpha$ in \cite{Ogata2006}), which is really  model dependent.  Secondly, in  \cite{Ogata2006},  the dependence upon  channel spin is ignored in the effective $p$-${\rm{^7Be}}$ interaction for the single particle wave function  of ${\rm{^8B}}$  and    the overestimated value of  $\lambda$(=0.159) was used (see, above and \cite{Tab06,Trache2003}).   Therefore,      under such  assumptions the $Z_B$ value extracted in \cite{Ogata2006}, which was used in Eq. (\ref{eq4}) together  with the single-particle ANC value for the adopted potential,      becomes strongly model dependent. For example,  the values of $\omega_{I}=\sqrt{Z_{B;\,I}/Z_B}$ in  terms of  which the calculated single particle cross section ($\sigma$) is parameterized (see, Eq. (45) in \cite{Ogata2006}),    are taken     equal to  0.397 and 0.918 for $I$=1 and 2, respectively, where $I$ is a channel spin. But, they are      equal to  0.497 and 0.868 for $I$=1 and 2 for   Barker`s spectroscopic factors ($Z_{B;\,I=1}$=0.251 and $Z_{B;\,I=2}$=0.765) \cite{Barker1983}. Apparently, all these facts are    the possible reasons why the ANC  derived in \cite{Ogata2006} is underestimated  in respect to that obtained in the present work.  In \cite{Trache01}, the  consideration is restricted only     by the Coulomb interactions    in  the  first-order perturbation  theory in the transition operator. However, as it was  demonstrated in \cite{Ogata2003}, this assumption is not correct and leads to    the underestimated value of the ANC obtained. Besides, in  \cite{Trache01} the three-body Coulomb postdecay acceleration effects in the final state of the  Coulomb  breakup reaction   is not taken into account. As  is shown quantitatively in \cite{Irgaz2005}, the influence of these effects grows  as the relative kinetic energy $E$ of the breakup fragments deceases, especially at extremely low energies.   But, our result for ANC obtained for $j_B$=3/2 is in  good  agreement with that of \cite{Belya2009} derived  from the ${\rm {^{58}Ni}}({\rm{^8B}}, p{\rm{^7Be}}){\rm{^{58}Ni}}$ analysis for the $1p_{3/2}$ proton orbital in ${\rm{^8B}}$. Besides,    the result of  \cite{Des2004a}, obtained within the three-body microscopic approach for the nucleon-nucleon MN potential, is aslo in a good agreement with that of the present work. It follows from here that the microscopic three-body ($\alpha {\rm{^3He}}\,p$) cluster calculation performed in \cite{Des2004a} for MN potential correctly reproduces the normalization of the tail of the radial overlap function of the ${\rm{^8B}}$ in the (${\rm{^7Be}}+p$)-channel.

\begin{figure}
\begin{center}
\includegraphics{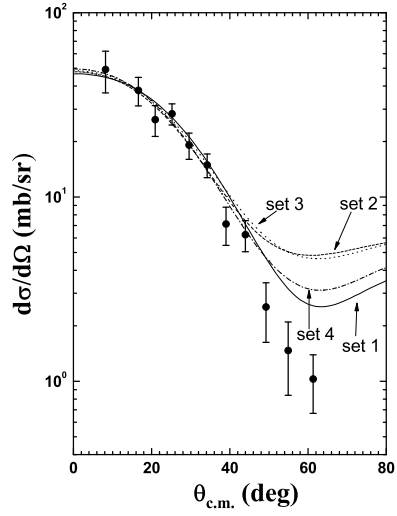}
\end{center}
\caption{\label{fig8} Angular distribution of the differential cross section   for the ${\rm{^7Be}}(d,n){\rm{^8B}}$ reaction at $E_i$=4.5   MeV   together with the theoretical calculations performed for the different sets of the optical parameters from Table \ref{table1} using the corresponding weighted  mean   of the ANC value  from Table \ref{table2}.  The experimental data are taken from Ref. \cite{Das2006}.   }
\end{figure}

The    weighted mean value for $C^2_B$ presented in Table \ref{table3} for the sets 1-4  of the  optical potentials   and the corresponding  central values of the ${\cal {R}}_{j_B}(E_i,\theta;\,b_{B;\,j_B})$ functions    were used in the expressions  (\ref{eq2}) and  (\ref{eq3}) for calculating the differential cross sections for the   ${\rm{^7Be}}$($d$,$n$)${\rm{^8B}}$  reaction at $E_i$=4.5  MeV.   The results of   calculations and their  comparison with the experimental data \cite{Das2006} are displayed in Fig. \ref{fig8}.  As  is seen from this figure,  the calculated cross section is in a good agreement with the experimental data in the main peak region of  the angular distribution for all  sets of the optical potentials.

 \begin{center}
{\bf III. APPLICATION FOR THE NUCLEAR ASTROPHYSICAL  REACTION AND THE EFFECTIVE-RANGE EXPANSION     }
\end{center}
 
 \begin{center}
{\bf C.    $S_{17}(E)$   for       ${\rm {^7Be}}(p,\gamma){\rm{^8B}}$ reaction  at solar energies } 
\end{center}

The ANC value      for $p+{\rm{^7Be}}\rightarrow{\rm{^8B}}$ presented in the last line of Table \ref{table3}  was used for calculation of  the astrophysical $S$ factor of the   ${\rm{^7Be}}(p,\gamma){\rm{^8B}}$ reaction at zero  energy  by using the formula \cite{Igam07}
\begin{equation}
\label{eq7}
S_{17}(0)={\rm 37.26}C^2_{{\rm{B}}}\,\, ({\rm{eV b}}).
\end{equation} 
  The obtained value of  $S_{17}(0)$  is  presented in Table \ref{table4} along with  the results   obtained within the other methods by  other authors.  As  is seen from Table \ref{table4}, the $S_{17}(0)$ value obtained in the present work  is in a good agreement  with that of     \cite{Igam08} and of \cite{Des2004a} obtained for the MN potential as well as with $S_{17}(0)$=23.27 eV b, which can be obtained from the interpolating  formula $S_{17}(E)$=23.27-40.53$E$+327.30$E^2$ derived by us from the polynomial formula (43) of  \cite{BB2000} for  Barker's potential and the above mentioned values    of the spectroscopic factors \cite{Barker1983}. At the same time, our result   differs  from  those of  \cite{Tab06,Ogata2006,Das2006,Ogata2003,Huang2010,Bar95,Baby2003,Baby2003a,Jungh2003}. In this connection, it should be noted the following. In \cite{Tab06,Ogata2006,Das2006,Ogata2003,Huang2010,Bar95} the value $S_{17}$(0)  was obtained by using   the   underestimated    value of the ANC in respect to that derived in the present work. In \cite{Ogata2006}, to obtain the value of   $S_{17}(0)$ the relation between   $S_{17}(0)/C_{B}^2$  and the $s$-wave ${\rm{^7Be}}+p$ scattering length \cite{Baye2000}  (see,  also below) was used in which the scattering lengths for $I$=1 and 2 were calculated with the Barker's potential. The ANCs  were extracted  there by using    the another form of the $p$-${\rm{^7Be}}$ potential by  ignoring its $I$-dependence, and  therefore, these calculations were not self-consistent. In \cite{Baby2003,Baby2003a,Jungh2003}, for obtaining  the $S_{17}$(0) value the  procedure of the    artificial fitting  the highly precise   experimental data measured there to the  astrophysical $S$ factors calculated in \cite{Des2004a}    was applied.  Therefore, the results of \cite{Baby2003,Baby2003a,Jungh2003} obtained for $S_{17}$(0) are model dependent since the results of \cite{Des2004a} presented also in Table \ref{table4} appreciably depend  on the form of the used  NN potentials.   The result of the solar fusion II  \cite{Adel2011}, which was complied also      from the results of \cite{Baby2003,Baby2003a,Jungh2003},  differs also noticeably on that obtained in the present work. Besides, as seen from Table \ref{table4}, there is the discrepancy between our result for $S_{17}$(0) and  those  of [48,58--61]  obtained from the $ {\rm{^8B}}$ Coulomb breakup analysis. Apparently, one of the possible reasons of this  discrepancy is the fact that, in [48,58--61],   first order over $\Delta V^{(C)}_f$  is used in the transition operator \cite{Ogata2006} and  the mentioned above   three-body Coulomb postdecay acceleration effects in the final state \cite{Irgaz2005} are not taken into account. Nevertheless, our result for $S_{17}(0)$ obtained only with taking into account the  value of ANC for $j_B$=3/2, which is equal to  20.6$\pm$0.73 eV b,  is in excellent  agreement with the result of \cite{Belya2009}. The latter has also been obtained  with taking into account  only  the   1$p_{3/2}$ orbital proton contribution in the ${\rm{^8B}}$ nucleus. 

  The results of extrapolation of the astrophysical $S$ factors at extremely low energies, including  close to the Gamov peak, $S_{17}$(20 keV),   $S_{17}$(50 keV) and  $S_{17}$(120 keV), which can be obtained within the MTBPA \cite{Igam07} by using the ANC values of the present work, are equal to  22.5$\pm$1.8,  21.7$\pm$1.8   and 20.8$\pm$1.5  eV b, respectively.
  
 We note that the ratio $S_{17}(E)/S_{17}(0)$ obtained from the results of the present work is equal to  0.97,    0.93 and 0.89    at $E$=20, 50 and 120 keV, respectively.  The same results for the ratio   can be obtained using the rational expression proposed in \cite{Jennis1998}. It follows from here that the ratio $S_{17}(E)/S_{17}(0)$ obtained from the results of the present work and   that of \cite{Jennis1998} reproduce correctly the energy dependence of  the $S_{17}(E)$ at solar energies ($E\lesssim$ 120 keV).

  Thus, the   $C_B^2$ value  and the values of  $S_{17}(E)$  at the solar energies ($E$=0, 20 and 50 keV)  derived     in the present work with the uncertainty about   7\% confirm the independent results of Ref. \cite{Igam08} obtained with the uncertainty about   3\%. Therefore, these  can   be considered as the ``best'' values  obtained by the indirect method so far. The $S_{17}(E)$ obtained in the present work and in \cite{Igam08} could  also be used as the main input data in Eq.(\ref{eq1}) for the correct estimation of the solar neutrino flux \cite{Bah1969,Adel2011}.   Nevertheless, more precise data are need for such  estimation of the boron neutrino flux. 
As it was shown in our paper, the experimental errors dominate in the ANC values extracted from analysis of the ${\rm{^7Be}}(d,n){\rm{^8B}}$ reaction at 4.5 MeV (c.m.). Moreover, this reaction becomes less peripheral at the larger energies.
So, it would be expedient to carry out new precise measurements of the ${\rm{^7Be}}(d,n){\rm{^8B}}$ reaction and the $d+{\rm{^7Be}}$ scattering at other near barrier energies of radioactive ${\rm{^7Be}}$ (less than 4.5 MeV (c.m.)) and as close to forward scattering angles as possible.

 \begin{center}
{\bf D.     $s$ wave $p+{\rm{^7Be}}$  scattering length   and the slop of $S_{17}(0)$}
\end{center}

  It is now of interest  to apply the ANCs and $S_{17}$(0) derived  by us above for obtaining information about  experimental values of the $s$ wave $p+{\rm{^7Be}}$  scattering length $a_o^{(I)}$ and their   average  one  $\bar{a}_o$ \cite{Baye2000} ($I$=1 and 2). To this end, we determine the squared  ANC values in other spin coupling modes by using the   relation $C^2_{B;\,I}=(C_{B;\,j_B=3/2}+(-1)^IC_{B;\,j_B=1/2})^2/2$ \cite{Blokh77} and the averaged values of the $C^2_{B;\,j_B}$   derived in the present work. They are equal to be $C^2_{B;\,I=1}$=0.116$\pm$0.014 fm$^{-1}$ and $C^2_{B;\,I=2}$=0.510$\pm$0.030 fm$^{-1}$, which results in $S_{17}^{(I)}(0)/C^2_{B;I}$= 3.685x10$^{-5}$ and 3.727x10$^{-5}$ MeV b fm obtained from (\ref{eq7}) for $I$=1 and 2, respectively.  From here and  formulas (27) and (28) of \cite{Baye2000}, we obtain $a_0^{(I)}\approx$23.2 and 14.8 fm for the $s$-wave scattering length for $I$=1 and 2, respectively, as well as the slope of  $S_{17}(0)$ near $E$=0 as $s_1^{(I)}=S_{17}{^{(I)}}^{\prime}(0)/S_{17}{^{(I)}}(0)\approx$-2.2 and -2.0   MeV$^{-1}$ for $I$=1 and 2, respectively. Using the values found  for  the   scattering lengths at $I$=1 and 2 in the formulas (29) and (30) of  \cite{Baye2000} we can obtain the values of the average scattering length $\bar{a}_o$ and then the value of  the slope $s_1=S_{17}^{\prime}(0)/S_{17}(0)$. They are    equal to be  $\bar{a}_o\approx$16.3 fm  and $s_1\approx$-2.1 MeV$^{-1}$. These results   for $s_1^{(I)}$ and $s_1$         are close to those of $s_1^{(1)}$=-1.65 MeV$^{-1}$,  $s_1^{(1)}$=-1.77 MeV$^{-1}$ and $s_1$=-1.74 MeV$^{-1}$,   which can be  obtained from the above mentioned polynomial approximation, as well as  to those of $s_1$=-1.86, -1.92  and -1. 97  MeV$^{-1}$ for the MN and V2 potentials \cite{Barker2006}, respectively.   
The slope  of  $S_{17}(0)$ near $E$=0  determined   in the present work becomes slightly steeper than that predicted in  \cite{Baye2000,Barker2006}. But,  as  is seen from here, the $s_1$ values of \cite{Baye2000,Barker2006}   depend noticeably on the input  potential. The values of $a_0^{(I)}$ ($I$=1 and 2) and $\bar{a}_o$ obtained in the present work  differ significantly  from those derived in \cite{Ogata2006,Baye2000,Angulo2003}.  In this connection, we   note that the magnitudes of $a_0^{(I)}$  and   $\bar{a}_o$ defined from the corresponding expressions of  \cite{Baye2000}  are very sensitive to those of the ratios $S_{17}^{(I)}(0)/C^2_{B;\,I}$  and  $S_{17}(0)/C^2_B$, where  $S_{17}(0)=\sum_{I=1,2}S_{17}^{(I)}(0)=\sum_{j_B=1/2,3/2}S_{17}^{(j_B)}(0)$  and $C^2_B=\sum_{I=1,2}C^2_{B;\,I}$. The calculations showed that a small change   of the ratios  results in   considerable  change for     $a_0^{(I)}$ and  $\bar{a}_o$.    For example, the values of  $a_0^{(I)}$=25 and -8 fm at $I$=1 and 2 as well as of  $\bar{a}_o$=-2.8 fm  obtained in  \cite{Ogata2006} give  the values of $S_{17}^{(I)}(0)/C^2_{B;\,I}$=3.677x10$^{-5}$ and 3.840x10$^{-5}$ MeV b fm for $I$=1 and 2, respectively, and $S_{17}(0)/C^2_B$=3.813x10$^{-5}$ MeV b fm.   The similar situation occurs in other works mentioned above. 

It follows from here that  one of the main reasons of the observed discrepancy between the results of the present work and other works for the  $s$-wave $p+{\rm{^7Be}}$ scattering lengths and the slope of $S_{17}(E)$ near $E$=0  is the  deference  between the    values of the ANCs  and $S_{17}$(0) derived by other authors and  those obtained in the present work. From our point of view, our results for   the  $s$-wave $p+{\rm{^7Be}}$ scattering lengths and the slope of $S_{17}(E)$ near $E$=0   are more trustable because they are derived  with the minimum ambiguity connected with the ANC and $S_{17}$(0) values.

\vspace{0.5cm}
\begin{center}
{\bf IV.   CONCLUSION}
\end{center}
\vspace{0.5cm}

\hspace{0.6cm} This scrupulous analysis of the  ${\rm{^7Be}}$($d$,$n$)${\rm{^8B}}$ reaction data at   $E_i$=4.5 and 5.8 MeV is performed  within the modified  DWBA. It is  demonstrated that  the  peripheral character of this reaction in the main peak  region of the angular distributions occurs  only for $E_i$=4.5 MeV \cite{Das2006}. Therefore, the experimental differential cross sections of the reaction under consideration measured in  \cite{Das2006}  can be used  as a source of determination of the squared ANC values $C^2_B$  for  $p+{\rm{^7Be}}\rightarrow{\rm{^8B}}$. A new value for the ANC was obtained, which is  in an   agreement with that recommended in \cite{Igam08,Belya2009} and differs strongly from the value, which is  deduced      within the modified  DWBA in \cite{Tab06} from the analysis of the other proton transfer reactions.  

The   value of the ANC from this work  was  used for  estimation of the astrophysical $S$ factor at $E$=0 and    the value  of $S_{17}(0)$   equal to  23.32$\pm$1.64 eVb was obtained, which is  in excellent  agreement with that recommended in  \cite{Igam08}. It  differs noticeably  from  that recommended in Refs . \cite{Adel2011,Tab06,Das2006,Huang2010,Bar95,Baby2003,Jungh2003} and obtained by  other authors from the data of the  $A(^8B,p^7Be)A$  Coulomb breakup reaction.    Also, the new estimation is obtained for the $s$-wave scattering length for the $p+{\rm{^7Be}}$ scattering and the slope  of  $S_{17}(0)$ near $E$=0.

\vspace{0.5cm}
\begin{center}
{\bf ACKNOWLEDGMENTS}
\end{center}
\vspace{0.5cm}

The authors are deeply grateful to L. D. Blokhintsev,  K. Ogata and R. J. Peterson for   careful reading of the manuscript, discussions  and constructive suggestions.   The work has been supported in part by the Academy of Sciences of the Republic of Uzbekistan (grants No. F2-FA-F117 and  F2-FA-F114).

 \newpage

\newpage

\newpage

\newpage

\newpage

\newpage

 \newpage

 {\selectlanguage{english}
 \begin{landscape}
\begin{table}[t]
 \caption{
\label{table1}  Parameters of the optical potentials corresponding to the entrance ($a$)  and exit ($b$) channels for the ${\rm{^7Be}}$($d$,$n$)${\rm{^8B}}$  reaction  at the energies $E_d$=5.79 and 7.46  MeV in the laboratory system.
 }
 \vspace{5mm}
\begin{tabular}{llllllllllll}
\hline
  \hline
 \multicolumn{1}{c}{$E_d$, }
  &\multicolumn{1}{c}{ set }
 &\multicolumn{1}{c}{ channel}
  &\multicolumn{1}{c} { $V$,   }
 &\multicolumn{1}{c}{$r_V$,  }
 &\multicolumn{1}{c}{$a_V$,  }
 &\multicolumn{1}{c}{4$W_D$($W$), }
 &\multicolumn{1}{c}{$r_D$($r_W$), }
&\multicolumn{1}{c}{$a_D$($a_W$), }
&\multicolumn{1}{c}{$V_{so}$, }
&\multicolumn{1}{c}{$r_{so}$, }
&\multicolumn{1}{c}{$a_{so}$, }\\
  \multicolumn{1}{c}{ (MeV)}
  &\multicolumn{1}{c}{  }
 &\multicolumn{1}{c}{ }
  &\multicolumn{1}{c} { (MeV)  }
 &\multicolumn{1}{c}{ (fm) }
&\multicolumn{1}{c}{  (fm)}
&\multicolumn{1}{c}{ (MeV)}
&\multicolumn{1}{c}{ (fm)}
&\multicolumn{1}{c}{ (fm)}
 &\multicolumn{1}{c}{ (MeV)}
&\multicolumn{1}{c}{ (fm)}
&\multicolumn{1}{c}{(fm)}\\
 \hline
5.79&1&\,\,\,\,\,$a$&72.0&1.05&0.95&(30.00)&(0.84)&(0.85)&12.8&1.05&0.94\\
&&\,\,\,\,\,$b$&28.0&1.64&1.05&121.2&1.94&0.11&4.9&1.64&0.27\\
&2&\,\,\,\,\,$a$&67.0&1.35&0.93&18.00&2.48&0.30&16.0&0.86&0.25\\
&&\,\,\,\,\,$b$&47.1&1.31&0.66&33.52&1.26&0.48&&\\
&3&\,\,\,\,\,$a$&66.2&1.35&0.93&17.43&2.43&0.30&16.0&0.86&0.25\\
&&\,\,\,\,\,$b$&35.8&1.51&0.43&25.98&1.84&0.51&5.8&1.18&0.51\\
&4&\,\,\,\,\,$a$&64.0&1.35&0.90&18.68&2.37&0.30&12.0&0.86&0.25\\
&&\,\,\,\,\,$b$&35.8&1.51&0.43&25.98&1.84&0.51&5.8&1.18&0.51\\
\hline
7.46&1$a$&\,\,\,\,\,$a$&65.0&1.35&0.89&19.16&2.34&0.30&12.0&0.86&0.25\\
 &&\,\,\,\,\,$b$&48.19&1.20&0.72&30.32&1.43&0.66&5.2&1.13&0.77\\
 &2$a$&\,\,\,\,\,$a$&64.0&1.35&0.90&18.68&2.37&0.30&12.0&0.86&0.25\\
&&\,\,\,\,\,$b$&48.2&1.13&0.72&45.32&1.43&0.66&6.2&1.13&0.77\\
&3$a$&\,\,\,\,\,$a$&64.0&1.35&0.90&18.68&2.37&0.30&12.0&0.86&0.25\\
&&\,\,\,\,\,$b$&42.4&1.35&0.55&37.56&1.35&0.75&5.0&1.35&0.55\\
\hline
\hline
\end{tabular}
 \end{table}
\end{landscape}

\newpage

  {\selectlanguage{english}
 \begin{table}[t]
\begin{center}
\caption{
\label{table2}   Coupling channel effects  for the ${\rm{^7Be}}$($d$,$n$)${\rm{^8B}}$  reaction  at $E_i$=4.5  MeV in the forward hemisphere calculated  for each   set of the optical potential used. Differential cross sections  and $\Delta_{CCE}$ are given in mb/sr and percentagewise, respectively. }
\vspace{5mm}
\begin{tabular}{lllll}
\hline
\hline
 \multicolumn{1}{c}{$\theta$,}&\multicolumn{4}{c}{$d\sigma_{CCE}/d\Omega$\,\,($\Delta_{CCE}$) }\\
 \cline{2-5}
  \multicolumn{1}{c}{(deg.)}
  &\multicolumn{1}{c}{ set 1}& \multicolumn{1}{c}{set 2}
   & \multicolumn{1}{c}{set 3}&\multicolumn{1}{c}{ set 4}\\
     \hline
     0.00&45.2(2.7)&34.96(2.9)&33.33(0.3)&34.60(1.1)\\
    8.20&44.6(3.2)&34.25(1.0)&32.60(1.0)&33.99(0.5)\\
    16.60&41.9(4.4)&31.30(1.0)&29.80(3.1)&31.51(1.3)\\
    20.90&39.2(5.5)&28.71(2.7)&27.38(6.4)&29.18(2.6)\\
    25.20&36.6(6.6)&25.80(4.3)&24.60(6.4)&26.47(4.0)\\
    29.60&26.4(9.9)&21.97(6.5)&20.90(8.5)&22.80(5.8)\\
    34.20&26.4(9.9)&17.85(8.8)&17.01(10.7)&18.84(7.7)\\
    \hline
    \hline
 \end{tabular}
\end{center}
 \end{table}

  \newpage

 {\selectlanguage{english}
 \begin{table}[t]
\begin{center}
\caption{
\label{table3} The dependence of  the weighted mean values of the squared  ANCs for ${\rm{^7Be}}+p\to{\rm{^8B}}$ from each of the sets 1--4 of the optical potentials  at $E_i$=4.5 MeV. Parenthetical figures are the weighted mean values of the squared  ANCs extracted without taking into account the CN and CCE contributions. }
 \vspace{5mm}
\begin{tabular}{lll}
\hline
  \hline
\multicolumn{1}{c|}{ set }&\multicolumn{1}{c}{$C^2_{B;\,3/2}$, fm$^{-1}$} &\multicolumn{1}{c}{$C^2_B$, fm$^{-1}$}\\
\hline
\,\,\,\,\,\,\,\,\,\,\,\,1&0.590$\pm$0.036&0.664$\pm$0.040\\
 &(0.563$\pm$0.034)&(0.633$\pm$0.039)\\
\,\,\,\,\,\,\,\,\,\,\,\,2&0.514$\pm$0.031&0.578$\pm$0.035\\
&(0.551$\pm$0.034)&(0.619$\pm$0.038)\\
\,\,\,\,\,\,\,\,\,\,\,\,3&0.587$\pm$0.035&0.660$\pm$0.039\\
&(0.641$\pm$0.038)&(0.721$\pm$0.042)\\
\,\,\,\,\,\,\,\,\,\,\,\,4&0.536$\pm$0.033&0.603$\pm$0.037\\
&(0.569$\pm$0.035)&(0.640$\pm$0.040)\\
 \hline
averaged  value&0.557$\pm$0.020(th)$\pm$0.034(exp)&0.626$\pm$0.022(th)$\pm$0.038(exp)\\
  &(0.581$\pm$0.021(th))$\pm$0.035(exp)&(0.653$\pm$0.024(theor)$\pm$0.032(exp))\\
  \hline
 &0.557$\pm$0.039&0.626$\pm$0.044\\ 
 &(0.587$\pm$0.041)&(0.653$\pm$0.046)\\
  \hline
 \end{tabular}
\vspace{-6mm}
\end{center}
\end{table}

\newpage

{\selectlanguage{english}
\begin{table}[h]
\begin{center}
\caption{\label{table4}{\small The squared ANC ($C^2_{B}= C^2_{B;\,1/2}+C^2_{B;\,3/2}$)
 for   ${\rm {^7Be}}+p\to\,\,{\rm {^8B}}$  and the astrophysical $S$ factor
($S_{17}$(0)) for the direct radiative capture ${\rm
{^7Be}}(p,\gamma){\rm {^8B}}$ reaction.}} \vspace{2 mm}
{\footnotesize
\begin{tabular}{llll}
 \hline
  \hline
  &&&\\
 \multicolumn{1}{c}{Method}&\multicolumn{1}{c}{$C^2_{B}$(fm$^{-1}$)}&\multicolumn{1}{c}{\hspace{-1.7cm}$S_{17}(0)$(eV$\cdot$b)}
&\multicolumn{1}{c}{\hspace{-1.0cm}Refs.}\\
&&&\\
 \hline
 &&&\\
 \hspace{0.6cm}  MDWBA&&& the  \\
 \hspace{0.5cm} ${\rm{^7Be}}(d,n){\rm{^8B}}$&0.626$\pm$0.022(th)$\pm$0.038(exp)&23.32$\pm$0.82(th)$\pm$1.42(exp)& present\\
  &0.626$\pm$0.044$^{1)}$&23.32$\pm$1.64$^{1)}$&work \\
  &&& \\
 \hspace{0.6cm}   MTBPA &&& \\
 \hspace{0.5cm} ${\rm{^7Be}}(p,\gamma){\rm{^8B}}$ &0.628$\pm$0.017&23.40$\pm$0.63& \cite{Igam08}\\
  &&& \\
 \hspace{0.6cm}MDWBA  &&&\\
  ${\rm {^{10}B}}({\rm{^7Be}},{\rm{^8B}}){\rm{^9Be}}$&&&\\
  ${\rm {^{14}N}}({\rm{^7Be}},{\rm{^8B}}){\rm{^{13}C}}$   &0.465$\pm$0.041&18.2$\pm$1.8&\cite{Tab06}\\
 \hspace{0.6cm}Breakup &&& \\
  ${\rm {^{208}Pb}}({\rm{^8B}},p{\rm{^7Be}}){\rm{^{208}Pb}}$&0.548&21.7$^{+0.37}_{-0.24}$(th)$\pm$0.50(exp)&\cite{Ogata2006}\\
   &&& \\
  \hspace{0.8cm}CDCCM  &&&\\
 \hspace{0.5cm}  ${\rm{^7Be}}(d,n){\rm{^8B}}$ &0.545$^{+0.036}_{-0.034}$(th)$\pm$0.070(exp)&20.96$^{+1.4}_{-1.3}$(th)$\pm$2.7(exp)& \cite{Ogata2003}\\
&&& \\ 
\hspace{0.8cm}R-matrix&&&\\
 \hspace{0.5cm}  ${\rm{^7Be}}(p,\gamma){\rm{^8B}}$ &0.518$^{2)}$&19.4$^{2)}$&\cite{Huang2010}\\
    &0.491&17.3$\pm$3.0&\cite{Bar95}\\
    &&& \\
  Coulomb breakup  &&& \\
\hspace{0.3cm}${\rm A} ({\rm{^8B}},p{\rm{^7Be}}){\rm A}$ &0.450$\pm$0.072&17.4$\pm$1.5&\cite{Trache01}\\
  ${\rm {^{58}Ni(^8B,p^7Be)^{58}Ni}}$&0.547$\pm$0.027$^{2)}$&20.8$\pm$1.1$^{2)}$&\cite{Belya2009}\\
&&& \\ 
&&& \\
  microscopic three- &0.812&29.45$^{3)}$  &\cite{Des2004a} \\
body ($\alpha{\rm {^3He}}p$) model& 0.668&24.65$^{4)}$&\cite{Des2004a}\\
&&& \\
 CDCCM DWBA &&& \\
 ${\rm{^7Be}}(d,n){\rm{^8B}}$ 
&&20.7$\pm$2.4&\cite{Das2006}\\
&&& \\
   phenomenological &&21.2$\pm$0.7&\cite{Baby2003,Baby2003a}\\
 \hspace{0.9cm} way &&21.4$\pm$0.6(th)$\pm$0.5(exp)&\cite{Jungh2003}\\
 &&& \\
    Coulomb breakup&&\\
${\rm {^{208}Pb}}({\rm{^8B}},p{\rm{^7Be}}){\rm{^{208}Pb}}$&&20.6$\pm$1.2$^{5)}\pm$1.0$^{6)}$&\cite{Iwasa1999}\\
 &&18.6$\pm$1.2$^{5)}\pm$1.0$^{6)}$&\cite{Schum2003a,Davids003}\\
&&20.6$\pm$0.8$^{7)}\pm$1.2$^{8)}$&\cite{Schum2003b}\\
&&& \\
  solar fusion II&&20.8$\pm$0.7$^{5)}\pm$ 1.4$^{6)}$&\cite{Adel2011}\\
\hline
\end{tabular}}
\\{\footnotesize
$^{1)}$ the overall uncertainty; $^{2)}$ for the $p_{3/2}$ state; $^{3)}$the V2 potential; $^{4)}$the MN potential;\\ $^{5)}$the exp. error; $^{6)}$the th. error; $^{7)}$ the stat. error; $^{8)}$the sys.error.}
 \end{center}
\vspace{-5.0mm}
\end{table}}

  \end{document}